\documentclass[aps,prb,twocolumn,superscriptaddress,floatfix,showpacs]{revtex4}
\usepackage{graphicx}
\usepackage{amssymb}
\usepackage{amsmath}
\usepackage{verbatim}
\usepackage{bbold}
\usepackage{booktabs}
\usepackage{bigstrut}
\usepackage{MnSymbol}

\begin{document}

\title{Optimal non-invasive measurement of Full Counting Statistics by a single qubit}

\author{A.V.\ Lebedev}
\affiliation{Theoretische Physik, Wolfgang-Pauli-Strasse 27, ETH Zurich,
CH-8093 Z\"urich, Switzerland}

\author{G.B.\ Lesovik}
\affiliation{L.D.\ Landau Institute for Theoretical Physics RAS,
Akad.\ Semenova av., 1-A, Chernogolovka, 142432, Moscow Region, Russia}

\author{G.\ Blatter}
\affiliation{Theoretische Physik, Wolfgang-Pauli-Strasse 27, ETH Zurich,
CH-8093 Z\"urich, Switzerland}

\date{\today}

\begin{abstract}
The complete characterisation of the charge transport in a mesoscopic device is provided by the Full Counting Statistics (FCS) $P_t(m)$, describing the amount of charge $Q = me$ transmitted during the time $t$.  Although numerous systems have been theoretically characterized by their FCS, the experimental measurement of the distribution function $P_t(m)$ or its moments $\langle Q^n \rangle$ are rare and often plagued by strong back-action. Here, we present a strategy for the measurement of the FCS, more specifically its characteristic function $\chi(\lambda)$ and moments $\langle Q^n \rangle$, by a qubit with a set of different couplings $\lambda_j$, $j = 1,\dots,k,\dots k+p$, $k = \lceil n/2 \rceil$, $p \geq 0$, to the mesoscopic conductor. The scheme involves multiple readings of Ramsey sequences at the different coupling strengths $\lambda_j$ and we find the optimal distribution for these couplings $\lambda_j$ as well as the optimal distribution $N_j$ of $N = \sum N_j$ measurements among the different couplings $\lambda_j$.  We determine the precision scaling for the moments $\langle Q^n \rangle$ with the number $N$ of invested resources and show that the standard quantum limit can be approached
when many additional couplings $p\gg 1$ are included in the measurement
scheme.
\end{abstract}

\pacs{
73.23.-b,  
05.60.Gg,  
06.20.-f,  
03.67.-a,  
}

\maketitle

\section{Introduction}

Traditionally, electronic transport through a device is characterized by the current and its noise. Within mesoscopic physics, Landauer's scattering matrix approach \cite{Landauer1957} provides a very physical and straightforward technique for the calculation of the average current
\cite{Landauer1957,Fisher1981,Imry1986, Buttiker1988} and noise
\cite{Lesovik1989,Buttiker1990,Martin1992,Blanter2000,Lesovik2011}, as well as higher moments.  The quantity which fully characterizes the random process of charge transport is given by the so-called Full Counting Statistics (FCS), telling what charge $Q = me$ is transmitted through the device during a fixed time $t$.\cite{noise_vs_FCS} The first calculation \cite{Levitov1992} of the probability distribution function $P_t(m)$ for the FCS goes back to 1992 and was quickly developed further \cite{LL1993,Ivanov1993,Levitov1996}. Various generalizations and applications have been proposed, e.g., the current noise in a normal-metal--superconductor point contact\cite{muzykantskii94}, the
electron transfer between superconductors\cite{belzig01}, charge pumping
\cite{andreev01} and charge transfer \cite{bagrets03} in the Coulomb blockade regime, the extension to energy-dependent scatterers \cite{hassler08}, the statistical properties of the persistent current in nanostructures \cite{komnik15}, or the fluctuations in the heat current in a quantum conductor \cite{utsumi14} or between two superconductors \cite{virtanen15}, to name just a few of the numerous theoretical studies.  At the same time, there are only very few experiments measuring higher-order correlators \cite{reulet03,bomze05,timofeev07} and one set of experiments measuring directly the  statistics\cite{lu03,schleser04,vandersypen04,gustavsson06}.
Unfortunately, measurement back-action is substantial in all of these
experiments and a non-invasive measurement of the Full Counting Statistics
remains to be done.  An early suggestion, formulated on the level of a
Gedankenexperiment and involving a spin \cite{Levitov1996} has later given way to a more concrete proposal based on charge- or flux-qubits \cite{hassler06}. However, a specific protocol how such a qubit is used in an optimal fashion is missing and it is the purpose of the present paper to close this gap.

The distribution function $P_t(m)$ and its moments or cumulants can be
obtained from the generating function $\chi_t(\lambda)$, the Fourier transform of $P_t(m)$, $\chi_t(\lambda) = \sum_m P_t(m) \exp(im\lambda)$. Here, $\lambda$ not only represents a compact variable in the Fourier transform---in its physical role it appears as the coupling constant between the transported charge and the qubit detector (here, we have in mind any qubit that couples to the charge either inductively or capacitively).  The basic quantity we are interested in then is the generating function $\chi(\lambda)$ and its derivatives with respect to $\lambda$. The latter define the moments (or cumulants) of the distribution function $P(m)$ (here and below we drop the time-index $t$ on $P_t(m)$ and $\chi_t(\lambda)$). The issue is to find the generating functions from measured data. This involves a simple protocol on the qubit with preparation, measurement, and a binary readout---the probabilities $P_{\pm}$ for the binary outcomes $+$ or $-$ at fixed $\lambda$ then allow for a statistical estimate $\tilde\chi(\lambda)$ of the generating function $\chi(\lambda)$. Evaluating $\tilde\chi(\lambda)$ for various values $\lambda = \lambda_j$ then allows for the determination of derivatives $\partial_\lambda^n \tilde \chi(\lambda)$ via finite difference formulas, from which estimates for the moments $\langle Q^n\rangle$ or cumulants can be obtained. The main question we want to answer in this paper then is: given a total of $N$ measurements, what is the optimal way to carry out these measurements? In particular, what number and distribution of grid points $\lambda_j$ shall be chosen, how should the $N$ measurements be distributed among the grid points, what accuracy can be achieved, and how does the precision of the result scale with the number $N$ of invested resources or measurements?

In Sec.\ \ref{sec:meas_chi} below, we will first describe the measurement
protocol providing estimates for the real and imaginary parts of the
characteristic function $\chi(\lambda)$ and analyze the statistical
distribution (or precision) of the measured results. The moments of
transferred charge involve higher-order derivatives of the characteristic
function $\chi(\lambda)$ and section \ref{sec:derivatives} is devoted to their construction out of measured values of $\chi$ through finite-difference formulas. The choice of grid-points in these finite-difference formulas interferes with the statistical errors from the measurements and one has to find the optimal grid and measurement strategy to minimize the total error for the charge moments; this task is discussed in Sec.\ \ref{sec:equi-points} for an equidistant set of coupling strengths and in Sec.\ \ref{sec:n-equi-points} for a non-equidistant set of grid points. The optimal measurement strategy involves a non-equidistant set of points and we find the optimal distribution of the number of measurements as well as the precision scaling with the total number $N$ of measurements. Specific results in the form of tables are given for the measurement of the third-order cumulant $\langle Q^3\rangle$. In section \ref{sec:conclusion} we present a summary, emphasize our main results, and add some concluding remarks on the use of different types of qubits and the relation to quantum counting \cite{lesovik10,suslov11}.

\section{Measurement of the characteristic function}\label{sec:meas_chi}

The Full Counting Statistics of a conductor can be described through the set of probabilities $P_t(m)$ to transmit $m$ particles (electrons) in a given
time $t$ (in the following we drop the index $t$). The discrete probability
distribution $P(m)$ can be characterised by a continuous generating function $\chi(\lambda) = \sum_m P(m) e^{im\lambda}$.  Given the generating function
$\chi(\lambda)$, one can find all moments of the transmitted charge (with the charge $Q$ measured in units of $e$),
\begin{equation}
   \mathcal{Q}_n \equiv \langle \hat{Q}^n \rangle
   = (-i)^n \lim_{\lambda \to 0} \partial_\lambda^n \chi(\lambda),
   \label{eq:moments}
\end{equation}
or the charge cumulants,
\begin{equation}
   \mathcal{K}_n \equiv \llangle \hat{Q}^n\rrangle
   = (-i)^n \lim_{\lambda \to 0} \partial^n_\lambda \ln\chi(\lambda).
\end{equation}
\begin{figure}[tb]
\includegraphics[width=7.5cm]{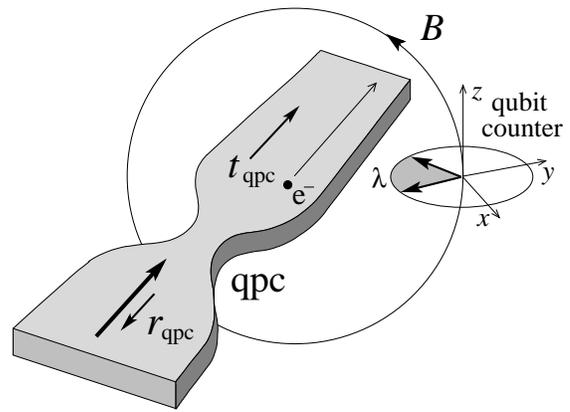}
\caption{Measurement of transmitted charge, e.g., across a quantum point
contact with transmission and reflection amplitudes $t_\mathrm{qpc}$ and
$r_\mathrm{qpc}$, respectively, by a qubit.  The passage of an electron
through the outgoing conductor generates a magnetic field pulse that rotates the qubit state (drawn as a vector on the Bloch sphere) by $\lambda$.}
\label{fig:setup}
\end{figure}

In order to find the above quantities in an experiment, we consider a qubit
locally interacting with the conductor as described by the Hamiltonian
$\hat{H}_\mathrm{int}= (\hbar\lambda/e) \hat\sigma_z \hat{I}(x,t)$, where
$\hat{I}(x,t)$ is the electric current operator in the conductor providing the transmitted charge at a position $x$ behind the scatterer, see Fig.\
\ref{fig:setup} for an illustration. Such a linear coupling is appropriate when the interaction point $x$ resides away from the scattering region in the conductor, see Bachmann {\it et al.} \cite{bachmann10}.  The qubit--current interaction leads to a rotation of the qubit state around the $z$-axis by an angle $\varphi = m\lambda$, where $m$ is the transmitted charge. Consider a standard Ramsey sequence of qubit rotations $\hat{R}(\varphi) = \hat{U}_y(-\pi/2) \hat{U}_z(\varphi) \hat{U}_y(\pi/2)$, where $\hat{U}_{\vec{n}}(\alpha) = \hat{\mathbb{1}} \cos(\alpha/2) -i \vec{n}\cdot \hat{\vec\sigma} \sin(\alpha/2)$, where the first and last rotations describe $\pm \pi/2$ rotations around the $y$-axis and the intermediate rotation is due to the interaction with the conductor.  Applying this Ramsey sequence to an initial qubit-state $|\! \uparrow\rangle$ (with $\sigma_z = 1$), one arrives at the final state
\begin{equation}\label{eq:n}
  |m\rangle = \hat{R}(\varphi = m\lambda)|\!\uparrow\rangle
   = \cos\frac{m\lambda}2 |\! \uparrow\rangle + i \sin\frac{m\lambda}2\,
   |\!\downarrow\rangle.
\end{equation}
The probabilities to observe the qubit in a state $\sigma_z = \pm 1$ then are given by
\begin{equation}\label{eq:Pnl}
   P_\pm(m\lambda) = \frac12 \pm \frac12 e^{-t/\tau_\varphi}\, \cos(m\lambda).
\end{equation}
The exponential damping in Eq.\ \eqref{eq:Pnl} accounts for the finite
dephasing time $\tau_\varphi$ of the qubit that we may model through a
stochastic Gaussian $H$-field. Such a finite dephasing time $\tau_\varphi$
ultimately limits the time $t$ during which the FCS $P_t(m)$ can be measured. For a particular run of the Ramsey sequence, the random number $m$ of transmitted charges is unknown but governed by the FCS distribution $P(m)$, hence, the probabilities for the two final qubit states can be found by averaging over $m$,
\begin{equation}\label{eq:Pl}
   P_\pm(\lambda) = \sum_m P(m)\, P_\pm(m\lambda).
\end{equation}
These probabilities are conveniently expressed through the real part of the
FCS characteristic function,
\begin{equation}\label{eq:spinprob}
   P_\pm(\lambda) = \frac12 \pm \frac12 e^{-t/\tau_\varphi} \mbox{Re}
   \chi(\lambda).
\end{equation}
Hence, repeating the Ramsey sequence $N \gg 1$ times and observing $N_\pm$
final outcomes with $\sigma_z = \pm 1$, one can directly estimate the real
part of the characteristic function for a given dimensionless
interaction parameter $\lambda$,
\begin{equation}
      e^{-t/\tau_\varphi}\mbox{Re}\tilde\chi(\lambda) = \frac{N_+ - N_-}{N_+ + N_-},
      \label{eq:reestimate}
\end{equation}
where the tilde $\tilde\chi(\lambda)$ refers to a statistical estimate of
$\chi(\lambda)$. The imaginary part of the characteristic function can be
estimated in a similar way by applying the alternative Ramsey sequence
$\hat{R}^\prime(\varphi) = \hat{U}_x(-\pi/2) \hat{U}_z(\varphi)
\hat{U}_x(\pi/2)$ to the initial state $\sigma_z = 1$ of the qubit.

In a realistic situation, the $\pi/2$ pulses in the Ramsey sequence are not
perfect, what modifies the statistical analysis of the data. Let us consider a non-perfect Ramsey sequence $\hat{R}(\varphi) = \hat{U}_y (-\pi/2 +
\delta^\prime) \hat{U}_z(\varphi) \hat{U}_y(\pi/2-\delta)$ with small
deviations $\delta$ and $\delta^\prime$. As a result, the perfect qubit
probabilities $P_\pm(\lambda)$ in Eq.\ (\ref{eq:spinprob}) are modified,
\begin{equation}
  P_\pm(\lambda) = \frac{1\pm\sin\delta\sin\delta^\prime}2 \pm
  \frac{e^{-t/\tau_\varphi}\cos\delta\cos\delta^\prime}2 \mbox{Re}
   \chi(\lambda).
\end{equation}
The imperfect $\pi/2$ pulses affect the result in two ways, i) an effective
decrease of the visibility factor, $e^{-t/\tau_\varphi} \to e^{-t/\tau_\varphi} \cos(\delta) \cos(\delta^\prime)$, which amounts to a renormalized dephasing time $\bar\tau_\varphi$ (at fixed $t$), and ii) a finite bias $P_+ - P_- = \sin(\delta)\sin(\delta^\prime)$ at $\lambda = 0$ that can be accounted for with a separate measurement. As a result, the estimated value of the characteristic function is given by
\begin{equation}
   e^{-t/\bar\tau_\varphi} \mbox{Re}\tilde\chi(\lambda)
   = \frac{N_+ - N_-}{N_+ + N_-}\bigg|_{\lambda}
   - \frac{N_+ - N_-}{N_+ + N_-}\bigg|_{0}.
\end{equation}
Note that the uncertainty in the dephasing time shows up in the final results with a small power $\alpha_{n,p}$, see, e.g., Eq.\ \eqref{eq:OPTL2}. In the following, we will assume perfect pulses with $\delta = 0 = \delta'$.

Next, we derive the statistical bounds for the estimation of $\tilde\chi$.
The experimental outcomes $N_\pm$ are distributed according to a
binomial distribution, $P(N_+,N_-) = C_N^{N_+} [P_+(\lambda)]^{N_+}
[P_-(\lambda)]^{N_-}$. As follows from Eq.\ (\ref{eq:spinprob}) this
distribution can be characterized by a single parameter $x =
\exp(-t/\tau_\varphi) \mbox{Re}\chi(\lambda)$ or $x = \exp(-t/\tau_\varphi)
\mbox{Im}\chi(\lambda)$. Hence, by virtue of Bayes theorem and observing a
particular set $N_\pm$ of results $\sigma_z = \pm 1$, one can obtain an
estimate of the {\it posterior} distribution function for the unknown
parameter $x \in [-1,1]$ via
\begin{equation}\label{eq:posterior}
   P(x| N_+, N_-) = \frac{(N+1)!}{2N_+! N_-!}\,
   \biggl( \frac{1+x}2\biggr)^{N_+} \biggl( \frac{1-x}2 \biggr)^{N_-}.
\end{equation}
For large $N_\pm$, the above distribution approaches a Gaussian, $P(x|N_+,N_-) \to {\cal N}(\tilde{x}, \sigma^2)$ with mean $\tilde{x} = [N_+-N_-]/N$ and variance $\sigma^2 = 4N_+N_-/N^3 = (1-\tilde{x}^2)/N$. Then, the statistical bounds for the estimated mean $x \approx \tilde{x}$ at a given tolerance level $\epsilon$ are given by,
\begin{equation}\label{eq:post-tol}
   \mbox{Prob}\Bigl[ |x - \tilde{x}| \leq g(\epsilon) \sigma \Bigr]
   = 1 - \epsilon,
\end{equation}
where $g(\epsilon)$ is determined by $1-\epsilon = \mbox{erf}(g/\sqrt{2})$ and the standard error function $\mbox{erf}(x) = (1/\sqrt\pi)\int_{-x}^x dt\, e^{-t^2}$, quickly approaching unity at large $x$, $\mbox{erf}(2) \approx 0.995$. Going back from the variable $x$ to the characteristic function $\chi(\lambda)$, one finds that with a probability $1-\epsilon$,
\begin{equation} \label{eq:chiest}
  |\mbox{Re}\chi(\lambda) - \mbox{Re}\tilde\chi(\lambda)| \leq g(\epsilon)\,
   \frac{v_\mathrm{Re}(\lambda)}{\sqrt{N}},
\end{equation}
where $v_\mathrm{Re}^2(\lambda) = \exp(2t/\tau_\varphi) - [\mbox{Re} \tilde
\chi(\lambda)]^2$ increases exponentially when pushing the measurement time
beyond $\tau_\varphi$. The same estimate holds true for the imaginary part of $\chi(\lambda)$.  The above measurement procedure then reaches the precision of the standard quantum limit at large $N$. We will see below that carrying over this standard quantum limit in the measurement precision for the moments $\mathcal{Q}_n$ is a non-trivial task [due to the appearance of derivatives $\partial_\lambda^n \chi(\lambda)$] and requires special measures.

Special attention has to be paid to the situation at small coupling $\lambda \gtrsim 0$, where $\mbox{Re}\chi(0) = 1$ and $\mbox{Im}\chi(0) = 0$. The
latter poses no problem as the distribution function Eq.\  (\ref{eq:posterior}) is centered around $x=0$, away from the boundaries at $x = \pm 1$, and hence well approximated by a Gaussian distribution with $v_\mathrm{Im}^2(\lambda) \leq v_\mathrm{Im}^2(0) \approx  e^{2t/\tau_\varphi}$.  On the contrary, when measuring the real part $\mbox{Re} \chi(\lambda \gtrsim 0)$, the distribution function (\ref{eq:posterior}) is squeezed towards the boundary at $x = 1$. In
this situation, $N_+ \sim N$ and (\ref{eq:posterior}) can be approximated by
\begin{equation}\label{eq:post-approx}
   P(x|N_+,N_-) = \frac{N_+^{N_-+1}}{2N_-!}
   \biggl(\frac{1-x}2\biggr)^{N_-} e^{-N_+(1-x)/2}.
\end{equation}
The maximum of (\ref{eq:post-approx}) is attained at $\tilde{x} = 1 - 2
N_-/N_+$ and provides an estimate for $x$ with an accuracy quantified by the variance $\sigma^2 = 4(N_-+1)/N_+^2$ and a precision scaling as $1/N$. With
increasing $N$, the number $N_-$ of outcomes $\sigma_z = -1$ increases and the distribution (\ref{eq:posterior}) detaches from $x = 1$ with
(\ref{eq:post-approx}) providing no longer a good approximation. Rather, the distribution (\ref{eq:posterior}) approaches the standard Gaussian form when $1-\tilde{x}$ becomes larger than $\sigma$, which is the case for $N_- \gg 1$ (using either of the above estimates for $\tilde{x}$ and $\sigma$).  In the following, we assume that $N$ is large enough, such that the Gaussian
approximation for the random variable $x = e^{-\tau/\tau_\varphi}
\mbox{Re}\chi(\lambda)$ provides a good description at any coupling strength $\lambda >0$.

\section{Calculation of Derivatives}\label{sec:derivatives}

Following Eq.\ (\ref{eq:moments}), the characteristic function $\chi(\lambda)$ (and its estimate $\tilde\chi$) can be used to determine the charge moments ${\cal Q}_n$.  This requires taking $n$-th order derivatives of $\chi(\lambda)$ near $\lambda = 0$, which can be found with the help of
finite-difference formulas of the form,
\begin{equation}\label{eq:fdf}
   \partial_\lambda^n \chi (\lambda)\big|_{\lambda = 0}
   \equiv \chi^{(n)} (0) \approx
   \sum_{\lambda\in\Lambda} w_\lambda^{(n)}\, \chi(\lambda),
\end{equation}
where $w_\lambda^{(n)}$ is a set of weight coefficients and $\Lambda =
\{\lambda_0, \lambda_1, \dots \lambda_m\}$ with $m \geq n$ is a set of
$\lambda$ values near the origin $\lambda = 0$. For a given $\Lambda$ and $n$, one can find the corresponding weight coefficients $w_\lambda$ (here and below we drop the index $^{(n)}$ on $w_\lambda^{(n)}$) using the
procedure described in Ref.\ \onlinecite{fornberg88}: defining $\omega(x)
\equiv \prod_{\lambda\in\Lambda}(x-\lambda)$, these are given as
\begin{equation} \label{eq:weight}
   w_\lambda = \frac{d^n}{dx^n} \frac{\omega(x)}{\omega^\prime(\lambda)
   (x-\lambda)} \biggr|_{x=0}
\end{equation}
with $\omega'(x) = \partial_x \omega(x)$. The characteristic function
$\chi(\lambda)$ is the Fourier transform of a real distribution function $P_n$ and hence $\chi(\lambda) = \chi^*(-\lambda)$. This symmetry motivates the use of symmetric sets $\Lambda_{n,p} =\{-\lambda_{k+p}, \dots,
-\lambda_1,0,\lambda_1, \dots, \lambda_{k+p}\}$, where $n = 2k$ or $n = 2k-1$ refer to even- or odd-order derivatives with $k>0$ and $p\geq 0$. This particular choice of the grid set $\Lambda_{n,p}$ twice reduces the number of points where $\chi(\lambda)$ has to be measured. Indeed, using Eq.\
(\ref{eq:weight}), one shows that $w_\lambda = w_{-\lambda}$ for even $n$
and $w_\lambda = -w_{-\lambda}$ when $n$ is odd. In addition, $\chi(0) = 1$
and one needs to measure $\chi(\lambda)$ only at the $k+p$ points $\lambda_j$ with $j=1,\dots,k+p$. Using the symmetry of the characteristic function $\chi(\lambda)$, the even (odd) derivatives of the characteristic function can be expressed through its real (imaginary) parts, respectively,
\begin{equation} \label{eq:findif}
   \chi^{(n)}(0) \approx \left\{
   \begin{array}{ll}
   \! w_0 + 2\sum\limits_{j=1}^{k+p} w_{\lambda_j} \mbox{Re}\chi(\lambda_j),
   & n = 2k,\\
   \! 2i\sum\limits_{j=1}^{k+p} w_{\lambda_j}\, \mbox{Im}\chi(\lambda_j),
   & n=2k-1.
   \end{array}
   \right.
\end{equation}
Making use of the Gaussian distributed estimates $\mbox{Re}\chi(\lambda_j)$
and $\mbox{Im}\chi(\lambda_j)$ characterized by Eq.\ (\ref{eq:chiest}) the
numerical derivatives Eq.\ (\ref{eq:findif}) are Gaussian random variables as well with a variance,
\begin{equation} \label{eq:sumvar}
   \mbox{Var}\bigl[\chi^{(n)}(0)\bigr]
   = 4g^2(\epsilon)\sum_{j=1}^{k+p} \, \frac{[w_{\lambda_j}
   v_n(\lambda_j)]^2} {N_j},
\end{equation}
where $v_n(\lambda)$ is equal to $v_{2k} = v_{\mathrm{Re}}$ ($v_{2k-1} =
v_{\mathrm{Im}}$) for even (odd) derivatives $n$ and $N_j$ is the number
of measurements that has been used to estimate the value of the characteristic function at $\lambda=\lambda_j$. Given a total number of measurements $N = \sum_j N_j$, the question poses itself how to distribute these resources over the $k+p$ measurement points. Minimizing $\mbox{Var} \bigl[ \chi^{(n)}(0) \bigr]$ under the condition of fixed $N$ one derives the following expression for the ratios $r_j \equiv N_j/N$ optimizing the distribution of measurements,
\begin{equation} \label{eq:nMeas}
   r_j = \frac{|w_{\lambda_j}| v_n(\lambda_j)}{\sum_{l=1}^{k+p}
   |w_{\lambda_l}| v_n(\lambda_l)},
\end{equation}
and the minimal variance is given by,
\begin{equation} \label{eq:staterr}
   \delta\mathcal{Q}_n^2 \equiv \mbox{Var}\bigl[ \chi^{(n)}(0)\bigr]
   = \frac{4g^2(\epsilon)}N \biggl[\sum_{j=1}^{k+p} |w_{\lambda_j}|
   v_n(\lambda_j)\biggr]^2.
\end{equation}

Having established the statistical error in the estimates of the derivatives of $\chi$, one also needs to take into account a second type of error arising due to approximation given by the finite difference formulas. E.g., choosing the grid points $\lambda\in \Lambda$ close to the origin $\lambda = 0$ decreases the error in the finite difference approximation (since the
remainder in the approximation (\ref{eq:fdf}) is of order  $\lambda^{n+p+2}$), however, the statistical error Eq.\ (\ref{eq:staterr}) grows due to the larger weights $w_\lambda \propto 1/\lambda^n$. Hence, we have to find the optimal grid $\Lambda_\mathrm{opt}$ that minimizes the total error given by the sum of statistical and approximation errors. This minimization introduces a dependence $\lambda_j(N)$ which will change (i.e., reduce) the overall precision scaling for the moments away from the standard quantum limit.

\subsection{Equidistant grids}\label{sec:equi-points}

Consider a measurement of the $n$-th moment of transferred charge ${\cal Q}_n$ by a set of $k+p$ qubits with equidistant coupling strengths $\lambda_j = j\lambda_0$, $j= 1, \dots, k+p$, where $n = 2k$ or $n = 2k-1$ and $p \geq 0$. Making use of Eq.\ (\ref{eq:weight}), the weights $w_{\lambda_j}$ in the finite difference formulas (\ref{eq:findif}) can be written in the form $w_{\lambda_j} = \kappa_j/\lambda_0^n$, where the coefficients $\kappa_j$ denote the set of numbers
\begin{equation} \label{eq:WEIGH2}
  \kappa_j = \frac{d^n}{dx^n} \frac{\omega(x)}
      {\omega^\prime(j)(x-j)} \bigg|_{x=0}
\end{equation}
with $\omega(x) = x\prod_{j=1}^{k+p}(x^2-j^2)$. Making use of Eq.\
(\ref{eq:sumvar}), the statistical error of the measurement then is given by
\begin{eqnarray} \label{eq:statERR}
   \delta {\cal Q}_n^2\Big|_\mathrm{stat}
   \nonumber
   &=& \frac{4g^2(\epsilon)}{\lambda_0^{2n}\, N}
   \sum_{j=1}^{k+p} \frac{\kappa_j^2}{r_j} \, v_n^2(j\lambda_0).
\end{eqnarray}
In the following, we approximate $v_{2k-1}(\lambda) \approx
e^{\tau/\tau_\varphi}$ and $v_{2k}(\lambda) \approx e^{\tau/\tau_\varphi} -
1$, hence, we assume that $v_n(\lambda)$ no longer depends on $\lambda$ near the origin; since $v_n(0) \leq v_n(\lambda)$, this corresponds to a conservative estimate of the statistical error.

The approximation error $\delta {\cal Q}_n|_{\mathrm{approx}}$ originating
from the finite difference formula approximating the derivative can be
obtained from Eq.\ (\ref{eq:findif}) by substituting the Taylor expansion
of $\mbox{Re}\chi(j\lambda_0)$ or $\mbox{ImRe}\chi(j\lambda_0)$; the first
$n+2p$ terms in this weighted (with the coefficients $w_{\lambda_j}$) sum
vanish (due to the very definition of the weights $w_{\lambda_j}$) and the
next term $\propto \lambda_j^{n+2p+2}$ provides an estimate for the
remainder
\begin{equation}
   \delta{\cal Q}_n\Big|_\mathrm{approx} = \lambda_0^{2p+2}\,
   |{\cal Q}_{n+2p+2}|\, \beta_{n,p},
\end{equation}
with the numerical (here, we introduce the coefficients $\nu_j = j$ for later reference, see Sec.\ \ref{sec:n-equi-points})
\begin{equation} \label{eq:beta}
   \beta_{n,p} = 2\frac{|\sum_{j=1}^{k+p} \nu_j^{n+2p+2}\kappa_j|}{(n+2p+2)!}.
\end{equation}
Minimizing the total error $\delta{\cal Q}_n = \delta {\cal
Q}_n|_\mathrm{stat} + \delta{\cal Q}_n|_\mathrm{approx}$ with respect to
$\lambda_0$, we find the minimal error
\begin{equation} \label{eq:MINERR}
   \delta{\cal Q}_n(\bar\lambda_0) = A_{n,p} \,
   |{\cal Q}_{n+2p+2}|^{1-2\alpha_{n,p}} \,\biggl[
   \frac{g^2(\epsilon) v_n^2}{N} \biggr]^{\alpha_{n,p}},
\end{equation}
with the scaling exponent
\begin{equation} \label{eq:alpha}
   \alpha_{n,p} = \frac{p+1}{n+2p+2}
\end{equation}
and the optimal distance $\bar\lambda_0$ between the couplings $\lambda_j$,
\begin{equation} \label{eq:OPTL}
   \bar\lambda_0 = B_{n,p}\, \biggl[\frac{g(\epsilon) v_n}
     {|{\cal Q}_{n+2p+2}|\sqrt{N}}\biggr]^{\alpha_{n,p}/(p+1)}.
\end{equation}
The exponent $1/(n+2) \leq \alpha_{n,p} < 1/2$ describes the precision
scaling of the experiment with the number $N$ of measurements.
The numericals $A_{n,p}$ and $B_{n,p}$ are given by the expressions
\begin{eqnarray} \label{eq:A}
   A_{n,p} &=& \frac{p+1}{n} \frac{\beta_{n,p}}{\alpha_{n,p}}
   \, [S_{n,p}]^{\alpha_{n,p}},\\
   \label{eq:B}
   B_{n,p} &=& [S_{n,p}]^{\alpha_{n,p}/2(p+1)},
\end{eqnarray}
with
\begin{equation} \label{eq:S}
   S_{n,p} = \frac{n^2 \sum_{j=1}^{k+p} \kappa_j^2/r_j}
                  {(p+1)^2\beta_{n,p}^2}.
\end{equation}

Finding the $n$-th order moment of the transmitted charge requires measuring $\mbox{Re}\chi(\lambda)$ or $\mbox{Im}\chi(\lambda)$ in at least at $k$
different values of the coupling constant $\lambda$. Analysing the scaling of the net error (\ref{eq:MINERR}) with respect to the number $N$ of
measurements, one notes that using only a minimal number of points, i.e.,
$p=0$, produces a small scaling exponent $\alpha_{n,0} = 1/(n+2)$ and hence
reaching a good precision implies a large number $N$ of measurements.  In
order to achieve a shorter overall duration of the experiment one needs to add more measurement points $p>0$; this strategy then allows to reach the standard quantum limit $\delta{\cal Q}_n \propto 1/\sqrt{N}$ at large $p$.
\begin{center}
\begin{table}[t] \caption{Weightfactors $\kappa_j$, scaling exponent
$\alpha_{3,p}$, and numericals $A_{3,p}$ and $B_{3,p}$ determining the
$3$-rd moment of transmitted charge for different additional grid points $p$.}
\label{tab:3-rd_order}\medskip
\begin{tabular}{|c|c|c|c|c|c|c|c|c|}
\hline
$p$&$\kappa_1$&$\kappa_2$&$\kappa_3$&$\kappa_4$ &$\kappa_5$&$\alpha_{3,p}$&
         $A_{3,p}$&$B_{3,p}$
\rule{0pt}{0ex} \rule[-1.25ex]{0pt}{0pt} \\
\hline
$0$&$-1$&$\frac12$ &$0$&$0$&$0$&$\frac15$&$1.32$&$1.78$
\rule{0pt}{2ex} \\
\midrule
$1$&$-\frac{13}8$&$1$&$-\frac18$&$0$&$0$&$\frac{2}{7}$&$1.55$&$1.84$
\rule{0pt}{2ex} \\
\midrule
$2$&$-\frac{61}{30}$&$\frac{169}{120}$&$-\frac{3}{10}$&$\frac{7}{240}$ &
         $0$&$\frac13$&$1.73$&$1.87$
\rule{0pt}{2ex} \\
\midrule
$3$&$-\frac{1669}{720}$&$\frac{4369}{2520}$&$-\frac{541}{1120}$&
       $\frac{1261}{15120}$&$-\frac{41}{6048}$&$\frac4{11}$&$1.89$&$1.89$
\rule{0pt}{2ex} \rule[-1.25ex]{0pt}{0pt} \\
\hline
\end{tabular}
\end{table}
\end{center}

Next, let us estimate the optimal coupling parameter $\bar\lambda_0$ as given by Eq.\ (\ref{eq:OPTL}). Assuming a driven (or non-equilibrium) charge transport, the higher moments scale as $|{\cal Q}_{n+2p+2}| \sim
|\bar{Q}|^{n+2p+2}$ with $\bar{Q} = \mathcal{Q}_1$ denoting the average
transmitted charge (in units of $e$). Then,
\begin{equation} \label{eq:OPTL2}
   \bar\lambda_0 \sim \frac{B_{n,p}}{|\bar{Q}|} \biggl[\frac{g(\epsilon) v_n}
   {\sqrt{N}}\biggr]^{\alpha_{n,p}/(p+1)},
\end{equation}
and the relative accuracy of the $n$-th moment $\delta{\cal Q}_n/|{\cal
Q}_n| \sim \delta{\cal Q}_n/|\bar{Q}|^n$ is given by,
\begin{equation} \label{eq:MINERR2}
   \frac{\delta{\cal Q}_n}{|\bar{Q}|^n} = A_{n,p}\,
   \biggl[\frac{g^2(\epsilon)v_n^2}{N}\biggr]^{\alpha_{n,p}}.
\end{equation}
Optimizing the proposed measurement scheme then requires a weak coupling
$\lambda$ between the conductor and the qubit, implying that the qubit is
typically rotated by an angle $\varphi \sim \bar{Q}\bar\lambda_0 \sim 1$ in
the $xy$-plane of the Bloch sphere during one Ramsey sequence (given the
smallness of the exponent, we drop the factor $N^{-\alpha_{n,p}/2(p+1)}$).
This result is quite natural, since at large couplings $\lambda$ the qubit
would perform multiple $2\pi$-rotations which cannot be distinguished by the proposed measurement scheme.
\begin{center}
\begin{table}[h] \caption{Relative number of measurements $r_j = N_j/N$ for
the $j$-th grid point.} \label{tab:3-rd_order_r}\medskip
\begin{tabular}{|c|c|c|c|c|c|}
\hline
$p$&$r_1$&$r_2$&$r_3$&$r_4$ &$r_5$
\rule{0pt}{0ex} \rule[-1.25ex]{0pt}{0pt} \\
\hline
$0$&$\frac23$&$\frac13$&$0$&$0$&$0$ \rule{0pt}{2ex} \\
\midrule
$1$&$0.59$&$0.36$&$0.05$&$0$&$0$\\
\midrule
$2$&$0.539$&$0.373$&$0.080$&$0.008$ &$0$\\
\midrule
$3$&$0.5012$&$0.3749$&$0.1044$& $0.0180$&$0.0015$\\
\hline
\end{tabular}
\end{table}
\end{center}

\subsubsection{Third-order charge moment}\label{sec:3rd}

Let us consider in more detail the measurement of the third-order charge
moment $\mathcal{Q}_3$ ($n=3$, $k=2$) for an equidistant grid with a different number of points $2 + p$, $p = 0, 1, \dots$. The corresponding weight factors in Eq.\ (\ref{eq:WEIGH2}), scaling exponent $\alpha_{3,p}$ and scaling factors $A_{3,p}$ and $B_{3,p}$ are presented in Table \ref{tab:3-rd_order} for $p=0,1,2,3$. Note that the numericals $A_{3,p}$ and $B_{3,p}$ are all of order unity.

The weights $\kappa_j$ in the finite-difference approximation assume higher
absolute values near the origin (small $j$) and are almost vanishing at large $j$. Therefore, most measurements have to be done for the first few grid points near the origin $\lambda= 0$; the relative number $r_j = N_j/ N$ of measurements [as they follow from Eq.\ (\ref{eq:nMeas})], are summarized in Table \ref{tab:3-rd_order_r} (the point $\lambda_0 =0$ requires no
measurement as $\chi(0) = 1$).
\begin{widetext}
\begin{center}
\begin{table}[h]
\caption{Optimal grid coefficients $\nu_j$, weightfactors $\kappa_j$, and
numericals $A_{3,p}$ and $B_{3,p}$ determining the $3$-rd moment of
transmitted charge for an additional number of grid points $p =
0,1,2,3$.}\label{tab:nu}\medskip
\begin{tabular}{|c|c|c|c|c|c|c|c|c|c|c|c|}
\hline
$p$&$\nu_2$&$\nu_3$&$\nu_4$&$\nu_5$&$\kappa_1$&$\kappa_2$& $\kappa_3$&
    $\kappa_4$&$\kappa_5$&$A_{3,p}$&$B_{3,p}$\\
\hline
$0$&$2.6180$&&&&$-0.5125$&$0.1957$&&&&$1.29$&$1.40$\\
\hline
$1$&$2.8019$&$4.0488$&&&$-0.6897$&$0.3182$&$-0.0499$&&&$1.46$&$1.36$\\
\hline
$2$&$2.8793$&$4.4113$&$5.4113$&&$-0.7676$&$0.3770$&$-0.0941$&$0.0180$&&
    $1.59$&$1.33$\\
\hline
$3$&$2.9188$&$4.6013$&$5.9109$&$6.7417$&$-0.8082$&$0.4087$&$-0.1208$&
    $0.0380$&$-0.0080$&$1.68$&$1.31$\\
\hline
\end{tabular}
\end{table}
\end{center}
\end{widetext}

\subsection{Non-equidistant grids}\label{sec:n-equi-points}

An equidistant set of grid points may not provide the optimal result, i.e.,
the smallest error $\delta \mathcal{Q}_n$. Hence, let us parametrize a
non-equidistant set of couplings $\Lambda = \lambda_0 \{1, \nu_2, \nu_3,
\dots, \nu_{k+p}\}$ by the minimal coupling $\lambda_0$ and an ordered set of $k+p$ constants $\nu_1 = 1 < \nu_2 < \dots < \nu_{k+p}$.  According to Eq.\ (\ref{eq:weight}), the finite-difference weights $w_{\lambda_j}$ have the form $w_{\lambda_j} = \kappa_j/\lambda_0^n$ where $\kappa_j$ can be found from Eq.\ (\ref{eq:WEIGH2}) with
\begin{equation}
      \omega(x) = x \prod_{j=1}^{k+p} (x^2 - \nu_j^2).
\end{equation}
Repeating the above analysis, one can minimize the sum of statistical and
approximation errors as a function of the coupling strength parameter
$\lambda_0$. The results (\ref{eq:MINERR}) and (\ref{eq:OPTL}) then hold true for the general grid $\Lambda$ with the replacement of the coefficients $\nu_j = j$ in Eq.\ (\ref{eq:beta}) by the distance coefficients $\nu_j$.
Dropping the requirement of equidistant grid points, one may attempt to
further optimize the factor $A_{n,p}$ in Eq.\ (\ref{eq:MINERR}) for given $n$ and $p$.  Although we have not been able to find an analytic expression for the coefficients $\nu_j$, we have performed a numerical optimization of
$A_{n,p}$ for $n=3$ and $p = 0,1,2,3$ as a function of $\nu_j$ with the
results shown in Table \ref{tab:nu} (the relative number $r_j$ of measurements remain those given in Table \ref{tab:3-rd_order_r}).

\subsection{Coupling strength sensitivity}\label{sec:lam_sens}

Another experimental limitation is due to imperfect knowledge of the coupling strengths $\lambda_j$. Assuming an accuracy $\delta\lambda_j$,
the weight coefficients $w_{\lambda_j}$ inherit an imprecision
\begin{equation}
   \delta w_{\lambda_j} = \sum_{l=1}^{k+p} \frac{\partial w_{\lambda_j}}
   {\partial\lambda_l}\, \delta\lambda_l
\end{equation}
and the resulting variation of the charge moment is given by
\begin{equation}\label{eq:dQ2}
   \delta{\cal Q}_n^2 = \sum_{l=1}^{k+p}
   \frac{\delta\lambda_l^2}{\lambda_l^2} \biggl[\sum_{j=1}^{k+p}
   \lambda_l\frac{\partial w_{\lambda_j}}{\partial \lambda_l}\,
   \mbox{Im}\chi(\lambda_j)\biggr]^2
\end{equation}
for the odd charge moments and a similar expression holds for the even
moments. A conservative estimate is obtained by replacing $|\chi(\lambda)|^2 \leq 1$ by unity in the above formula.  The derivatives $\lambda_l
\partial_{\lambda_l} w_{\lambda_j} = (\nu_l/\lambda_0^n)
\partial_{\nu_l} \kappa_j$ can be found (numerically) from Eq.\
(\ref{eq:WEIGH2}) for a given set $\Lambda$ of coupling strengths. For
simplicity, we assume that all coupling parameters $\lambda_j$ are known with the same relative accuracy $\epsilon_\lambda = \delta\lambda_j/\lambda_j$, then
\begin{equation}
   \delta{\cal Q}_n^2 \leq \frac{E_{n,p}^2}{\bar\lambda_0^{2n}} \,
   \epsilon_\lambda^2,
\end{equation}
where we have replaced $\lambda_0$ by $\bar\lambda_0$, see Eq.\
(\ref{eq:OPTL2}), and
\begin{equation}
   E_{n,p}^2 = \sum_{l = 1}^{k+p}\,\, \biggl[\sum_{j=1}^{k+p}
   \nu_l \frac{\partial\kappa_j}{\partial\nu_l}\biggr]^2
\end{equation}
is a numerical factor which depends only on the set of relative coupling
strengths $\nu_j$. Given a relative accuracy $\epsilon_\lambda$, one can find the total number of measurements $\bar{N}$ required to give the same measurement precision $\delta{\cal Q}_n$ as in (\ref{eq:MINERR2}),
\begin{equation}
   \bar{N} \sim \biggl[g(\epsilon) \, v_\mathrm{Im}
   \frac{A_{n,p} B_{n,p}^n}{E_{n,p}}\biggr]^2 \frac1{\epsilon_\lambda^2}.
\end{equation}
A further increase of $N$ beyond $\bar{N}$ does not improve the precision of $\mathcal{Q}_n$. In Table \ref{tab:S} below, we list the corresponding factors $E_{3,p}$ for the measurement of the third-order charge moment with
an equidistant and an optimal grid.
\begin{center}
\begin{table}[h]
\caption{Coefficients $E_{3,p}$ quantifying the accuracy $\delta \mathcal{Q}_3$
under uniform variation $\epsilon_\lambda = \delta \lambda_j/\lambda_j$ of the
couplings $\lambda_j$.}\label{tab:S}\medskip
\begin{tabular}{|c|c|c|c|c|}
\hline
$\Lambda$&$S_{3,0}$&$S_{3,1}$&$S_{3,2}$&$S_{3,3}$\\
\hline
$\Lambda^\mathrm{eq}$&$1.12$&$1.91$&$2.49$&$2.94$\\
\hline
$\Lambda^\mathrm{opt}$&$0.55$&$0.76$&$0.86$&$0.91$\\
\hline
\end{tabular}
\end{table}
\end{center}
Note that the measurement involving the optimal grid is less sensitive to the errors in $\lambda_j$ as compared with the measurement based on the equidistant grid and requires less measurements $\bar{N}$. For example, the measurement with $p=3$ provides a scaling exponent $\alpha_{3,3} = 4/11$ and using a non-equidistant optimal grid one arrives at the results
\begin{equation}\label{eq:opt_grid}
   \bar{N} \approx 17.2 \frac{g^2(\epsilon) v_\mathrm{Im}^2}{\epsilon_\lambda^2},
   \quad \frac{\delta{\cal Q}_3}{|\bar{Q}|^3} \sim 0.21\, \epsilon_\lambda^{8/11},
\end{equation}
while using an equidistant grid leads to values
\begin{equation}\label{eq:equi_grid}
   \bar{N} \approx 18.8 \frac{g^2(\epsilon) v_\mathrm{Im}^2}{\epsilon_\lambda^2},
   \quad \frac{\delta{\cal Q}_3}{|\bar{Q}|^3} \sim 0.22 \,\epsilon_\lambda^{8/11}.
\end{equation}

\subsection{Charge cumulants}\label{sec:charge_cumulants}

In addition to the charge moments one might be interested in the charge
cumulants ${\cal K}_n = \llangle \hat{Q}^n \rrangle$; the latter can be
expressed through a combination of charge moments ${\cal Q}_m$ with
$m \leq  n$, for example,
\begin{equation}
      {\cal K}_3 ={\cal Q}_3 - 3 {\cal Q}_2 {\cal Q}_1 + 2 {\cal Q}_1^3.
\end{equation}
Given a grid $\Lambda_{n,p} = \{\lambda_1, \dots, \lambda_{k+p}\}$ of $k+p$
measuring points, one has to measure all charge moments ${\cal Q}_m$ with
$m\leq n$. The most imprecise measurement in the above combination is given by the measurement of the highest charge moment ${\cal Q}_n$, hence, this
measurement has to be fully optimized with respect to the number of
measurement proportions $r_j$ as well as optimal coupling strengths
$\lambda_j$. As follows from the Eq.\ (\ref{eq:OPTL2}), the optimal
value of $\lambda_0$ for each $\mathcal{Q}_m$ measurement with $m \leq n$ is given by
\begin{equation}
   \bar\lambda_0^{(m)} \sim B_{m,p+n-m}|\bar{Q}|^{-1}\,
   \biggl[ \frac{g(\epsilon) v_m}{\sqrt{N}}\biggr]^{1/(2n+2p-m+2)}.
\end{equation}
The precision scaling in $N$ involves only small exponents $1/(2n+2p-m+2)$ and thus all couplings $\bar\lambda_0^{(m)}$ are of the same order as
$\bar\lambda_0^{(n)}$.  The lower charge moments with $m < n$ then can be
measured using the same grid $\Lambda_{n,p}$ and hence the same data
$\tilde\chi(\lambda_j)$ with different weights $w_{\lambda_j}^{(n)}$. Finally, as the main contribution to the measurement error of ${\cal K}_n$ originates from the measurement error of the largest charge moment, we have
\begin{equation}
      \delta{\cal K}_n \sim \delta{\cal Q}_n
\end{equation}
with an optimized $\delta\mathcal{Q}_n$.

\section{Conclusion}\label{sec:conclusion}

In this paper, we have derived an optimized strategy for measuring the charge moments $\mathcal{Q}_n = \langle \hat{Q}^n\rangle$ (and hence the Full Counting Statistics) in the random charge transfer across a mesoscopic device. These moments appear as derivatives of the generating function $\chi(\lambda)$, which can be measured with the help of a qubit performing Ramsey sequences at couplings $\lambda$. The derivatives of $\chi(\lambda)$ can be determined with the help of finite-difference formulas involving a set of measurements at different couplings $\lambda_j$, $j = 1, \dots k+p$, with a minimal number of couplings $k = \lceil n/2 \rceil$ and $p \geq 0$ additional grid-points. Given a total number $N$ of Ramsey sequences, we have found the optimal distribution $N_j$ of such measurements among the different couplings $\lambda_j$.  For an equidistant grid, we have found the optimal grid separation $\bar{\lambda}_0$ and the exponent $\alpha$ of the precision scaling, $\delta \mathcal{Q}_n \propto N^{-\alpha}$. The typical coupling $\bar{\lambda}_0$ then generates a rotation $\varphi \sim 2\pi$ on the passage of the average charge $\bar{Q}$ during a Ramsey sequence, $\bar{Q} \bar{\lambda}_0 \sim \varphi/2\pi \sim 1$.  The precision exponent
$\alpha_{n,p}$ depends on the order $n$ of the moment and the number $p$ of
additional grid points. Higher moments come with a poor scaling $\alpha_{n,0} = 1/(n+2)$ for the minimal grid with $p=0$. On the other hand, choosing a large $p$ is beneficial and allows to approach the standard quantum limit $\alpha_\mathrm{sql} = 1/2$. The best set of couplings $\lambda_j$ is not equidistant; unfortunately, finding this grid requires a numerical optimization. Such an optimization, as well as the determination of all other relevant quantities and numericals, has been done for the measurement of the third-order moment $\mathcal{Q}_3$. Another requirement is the precise knowledge of the couplings $\lambda_j$, as a relative imprecision $\epsilon_\lambda$ of the couplings $\lambda_j$ limits the number $N$ of useful measurements to a value $\bar{N} \propto 1/\epsilon_\lambda^2$. An interesting observation is that the non-equidistant optimized grid provides a better precision with fewer measurements as compared with the equidistant grid, although the difference is small, see Eqs.\ (\ref{eq:opt_grid}) and (\ref{eq:equi_grid}).

Let us also discuss a few more subtle issues related to the measurement of the full counting statistics.  First, we point out that the measured probability function or correlators depends on the type of qubit. E.g., a flux qubit measures the passage of directed charge (plus for right-moving, minus for left-moving) and hence quantifies the statistics of the net transferred charge. On the other hand, a charge qubit accumulates the signal from passing charge independent of its direction of motion and hence provides a characterization of the total charge transferred across the detector (in any direction); both quantities are measured perfectly fine with the above recipe. It is then the experimenter who has to decide about the appropriate type of qubit that measures the quantity of interest.

A second issue is the location of the qubit detector, close or far away from the scattering region at $x=0$.  This question relates to some fundamental
concerns that appeared very early on in the context of extending the transport characteristic beyond the noise correlator.  In fact, different results have been obtained for the 3-rd order cumulant, once the quantum binomial expression \cite{Levitov1992} $\llangle Q^3 \rrangle_q \propto -2T^2(1-T)$ when no time-ordering was imposed or when placing the qubit near the scatterer \cite{bayandin08}, while the classical binomial result
\cite{LL1993,Levitov1996} $\llangle Q^3 \rrangle_c \propto T(1-T)(1-2T)$ has been found when time ordering was applied as prescribed through the inclusion of a spin detector into the description. The problem has been resolved recently \cite{bachmann10} with the demonstration that the classical binomial result applies everywhere, far away as well as close to the scatterer, at least for the case of a spin or flux qubit device.

Another remark concerns the relation of measuring the FCS to the problem of
quantum counting \cite{lesovik10,suslov11}, where a qubit register of $K$
qubits with coupling strengths $\lambda_j  = \pi/2^{j-1}$, $j = 1,\dots,K$ can be used to find the precise number of particles $m < 2^K$ that has been
transmitted during a time $t$. Repeating the measurement many times and
averaging then allows to find the FCS $P_t(m)$ as well. However, in this case, the strongest coupling $\lambda_1 = \pi$ rotates the qubit $j=1$ by $\pi$ on passage of a single electron, so much stronger couplings are required than in the present protocol where $\lambda_0 \sim 1/\bar{Q}$. Furthermore, quantum counting and subsequent averaging provides much more information than needed if the goal is the measurement of a cumulant $\mathcal{Q}_n$.

\acknowledgments

We acknowledge financial support from the Swiss National Science Foundation
through the National Center of Competence in Research on Quantum Science and Technology (QSIT), the Pauli Center for Theoretical Studies at ETH Zurich, and the RFBR Grant No.\ 14-02-01287.


\begin{thebibliography}{99}
\bibitem{noise_vs_FCS} Note that the measurement of FCS differs from the
measurement of current correlators, e.g., current noise or higher order
correlators. The latter involves time-resolved or finite frequency
measurements. The FCS relates to the system's low-frequency properties, e.g., zero-frequency noise, in the long time limit.
\end{thebibliography}
\end{document}